\documentclass[9pt,twocolumn,twoside]{optica}
\setboolean{shortarticle}{false}
\setboolean{minireview}{false}

\newcommand{\CMT}[1]{{}}

\title{Waveguide cavity optomagnonics for broadband multimode microwave-to-optics conversion}

\author[1]{NA ZHU}
\author[1]{XUFENG ZHANG}
\author[1]{XU HAN}
\author[1]{CHANG-LING ZOU}
\author[2]{CHANGCHUN ZHONG}
\author[2]{CHIAO-HSUAN WANG}
\author[2]{LIANG JIANG}
\author[1,*]{HONG X. TANG}

\affil[1]{Department of Electrical Engineering, Yale University, New Haven, Connecticut 06511, USA }
\affil[2]{Department of Applied Physics, Yale University, New Haven, Connecticut 06511, USA}

\affil[*]{Corresponding author:  hong.tang@yale.edu}

\begin{abstract}
Cavity optomagnonics has emerged as a promising platform for studying coherent photon-spin interactions as well as tunable microwave-to-optical conversion. However, current implementation of cavity optomagnonics in ferrimagnetic crystals remains orders of magnitude larger in volume than state-of-the-art cavity optomechanical devices, resulting in very limited magneto-optical interaction strength. Here, we demonstrate a cavity optomagnonic device based on integrated waveguides and its application for microwave-to-optical conversion. By designing a ferrimagnetic rib waveguide to support multiple magnon modes with maximal mode overlap to the optical field, we realize a high magneto-optical cooperativity which is three orders of magnitude higher compared to previous records obtained on polished YIG spheres. Furthermore, we achieve tunable conversion of microwave photons at around 8.45\,GHz to 1550\,nm light with a broad conversion bandwidth as large as 16.1 MHz. The unique features of the system point to novel applications at the crossroad between quantum optics and magnonics.

\end{abstract}

\setboolean{displaycopyright}{true}

\begin{document}

\maketitle

\section{Introduction}

Exploitation of hybrid platforms to combine microwave and photonic circuits is of great interest for its importance to realize both classical and quantum hybrid signal transduction, storage, and processing \cite{kimble2008quantum,xiang2013hybrid,schoelkopf2008wiring,o2009photonic,lvovsky2009optical}. Coherent microwave-to-optical conversion has been realized at various experimental platforms, for example, by coupling with auxiliary excitations such as phonons in optomechanical systems \cite{bochmann2013nanomechanical,andrews2014bidirectional,balram2016coherent,lecocq2016mechanically,fan2016integrated,fan2015cascaded,shao2019microwave,han2020cavity}, and direct microwave-light interaction via the electro-optic approach \cite{tsang2010cavity,rueda2016efficient,fan2018superconducting}. While benefiting from the small mode volume to enhance the interactions, most systems intrinsically lack tunability and have relatively narrow operating bandwidths ($<$ 1 MHz), limiting their practical applications.
In recent years, the coherent, cavity-enhanced interaction between optical photons and solid state magnons has attracted intensive attentions in both fundamental research and device applications, because of the appealing properties of magnons such as long spin lifetime and large-bandwidth tunability. In particular, single crystalline ferrimagnetic insulator yttrium iron garnet (YIG, Y$_{\text{3}}$Fe$_{\text{5}}$O$_{\text{12}}$) has emerged as a promising candidate for integrating magnons in hybrid quantum systems to bridge different types of excitations. It has been widely adopted to investigate interactions among spin waves, microwaves, acoustic waves, and optical excitations \cite{lachance2019hybrid,hisatomi2016bidirectional,osada2016cavity,osada2018brillouin,haigh2016triple,kusminskiy2016coupled,zhang2015magnon,zhang2016cavity,zhang2014strongly,zhang2016optomagnonic,li2019strong,harder2018level,rao2019analogue}. YIG exhibits very low dissipation for all these information carriers and in particular, as an optical material, it shows very low optical loss in the telecom c-band (0.13 dB/cm) \cite{wood1967effect}. Because of these advantages, the feasibility of using single crystalline YIG for realizing microwave-to-optical conversion is of significant interest \cite{kusminskiy2016coupled,graf2018cavity,sharma2019optimal,pantazopoulos2019high,kusminskiy2019cavity,haigh2018selection,vsimic2020coherent}. However, previous studies in this area mainly focused on bulk crystals with the uniform magnon mode (Kittel mode) \cite{kittel1958excitation}, resulting in the low magneto-optical interaction strength and limited conversion bandwidth \cite{liu2016optomagnonics,zhang2016optomagnonic,hisatomi2016bidirectional,osada2016cavity,osada2018brillouin,haigh2016triple}.

\begin{figure*}[htbp]
\centering
\includegraphics[width= 155 mm]{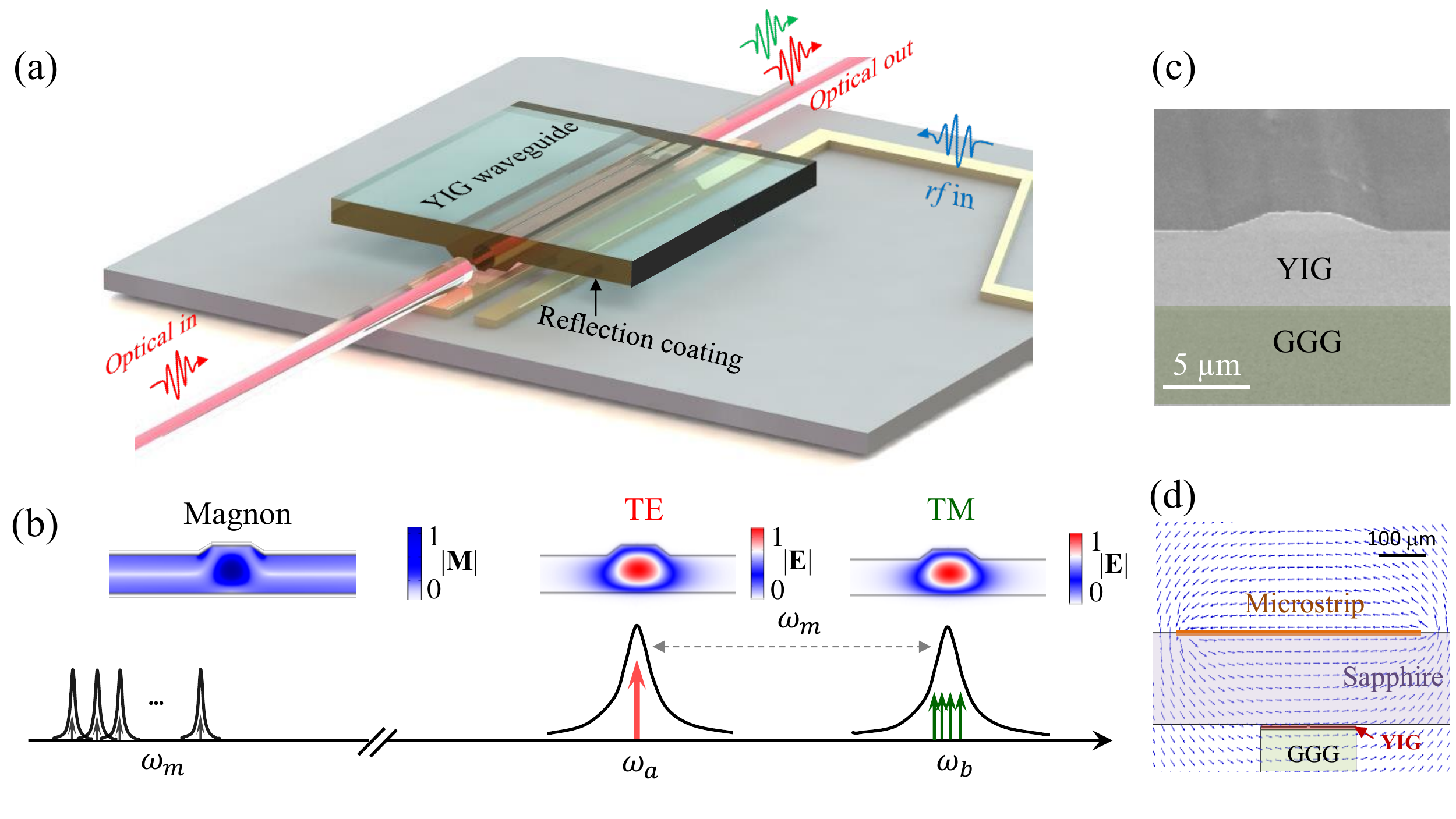}
\caption{(a) Schematic of the experimental assembly of the integrated optomagnonic waveguide device. The optical light is coupled in and out of the YIG rib waveguide via cleaved fibers. A \textit{rf} microstrip cavity is aligned beneath the waveguide for magnon excitations.  (b) A frequency domain representation of the triple-resonance-enhanced frequency conversion process. A strong optical pump light is applied to the TE optical mode (pump mode). Photons can be converted between the magnon modes and TM optical mode (signal mode). The electric field distributions in the cross-section view for optical TE, TM, and the magnetic field distribution of the magnon modes are shown in the upper panel, respectively. (c) False-color scanning electron microscope image of the waveguide cross-section view. (d) The \emph{rf} magnetic field distribution across the YIG waveguide at the cross-section view. 
}
\label{figure1}
\end{figure*}


In this paper, we experimentally demonstrate the multimode-magnon assisted, broadband, and tunable conversion between microwave and optical light in a single crystalline ferrimagnetic YIG thin film waveguide configuration. The waveguide simultaneously supports a series of Fabry-P\'erot (FP) optical resonances and multiple magnon resonances formed by forward volume magnetostatic standing waves (FVMSWs) \cite{stancil2009spin}. With the carefully designed YIG rib-waveguide geometry, the magnon modes and optical modes are both confined in a small mode volume with large mode overlap, which results in significantly enhanced magneto-optical interactions compared with previous studies in YIG spheres \cite{hisatomi2016bidirectional,zhang2016optomagnonic,osada2016cavity}. At the same time, a second optical cavity mode is engineered to resonantly boost the intra-cavity pump photon number. Taking advantage of the triple-resonance integration by realizing the energy and phase conservation among magnon and optical photons, we demonstrate a microwave-to-optical photon conversion efficiency that is three orders of magnitude higher compared with current state-of-the-art results on YIG-based platforms \cite{hisatomi2016bidirectional,zhang2016optomagnonic,PhysRevLett.121.199901,osada2016cavity}. Furthermore, with the existence of multiple magnon resonances, a broad frequency conversion bandwidth exceeding 16.1 MHz has been realized. Our demonstration sheds light on the potential of the patterned YIG rib waveguide as a new platform for magnon-based coherent information processing.

\section{Framework of waveguide cavity optomagnonics mediated conversion}

The architecture of the optomagnonic device is illustrated in Fig. \ref{figure1}(a). The system consists of multiple coupled modes in three different domains: a microwave cavity mode supported by a half-$\lambda$ resonator, magnon modes formed by forward volume magnetostatic standing waves in an etched YIG waveguide, and optical modes engineered to have the desired dispersion relation for supporting the triple-resonance enhanced frequency conversion, which means that the magnon, input and output optical photons are simultaneously on resonance. The magnons are coupled to the microwave cavity which is driven by the input itinerant microwave photons through an inductively coupled microwave feedline. Due to the coexistence of the magnonic and the optical cavity modes and the strong magneto-optical interaction, input optical photons can be inelastically scattered by the magnons into a single sideband mode with orthogonal polarization via the \emph{Faraday effect} \cite{van1965optically,zhang2016optomagnonic,hisatomi2016bidirectional,haigh2016triple}, with the frequency difference matching the magnon mode frequency. This can be described by an intuitive picture that, under the microwave drive, the polarization of the pump light oscillates at the frequency of the magnons, and thus produce the desired optical sideband under phase matching conditions.

\begin{figure*}[htbp]
\centering
\includegraphics[width= 175 mm]{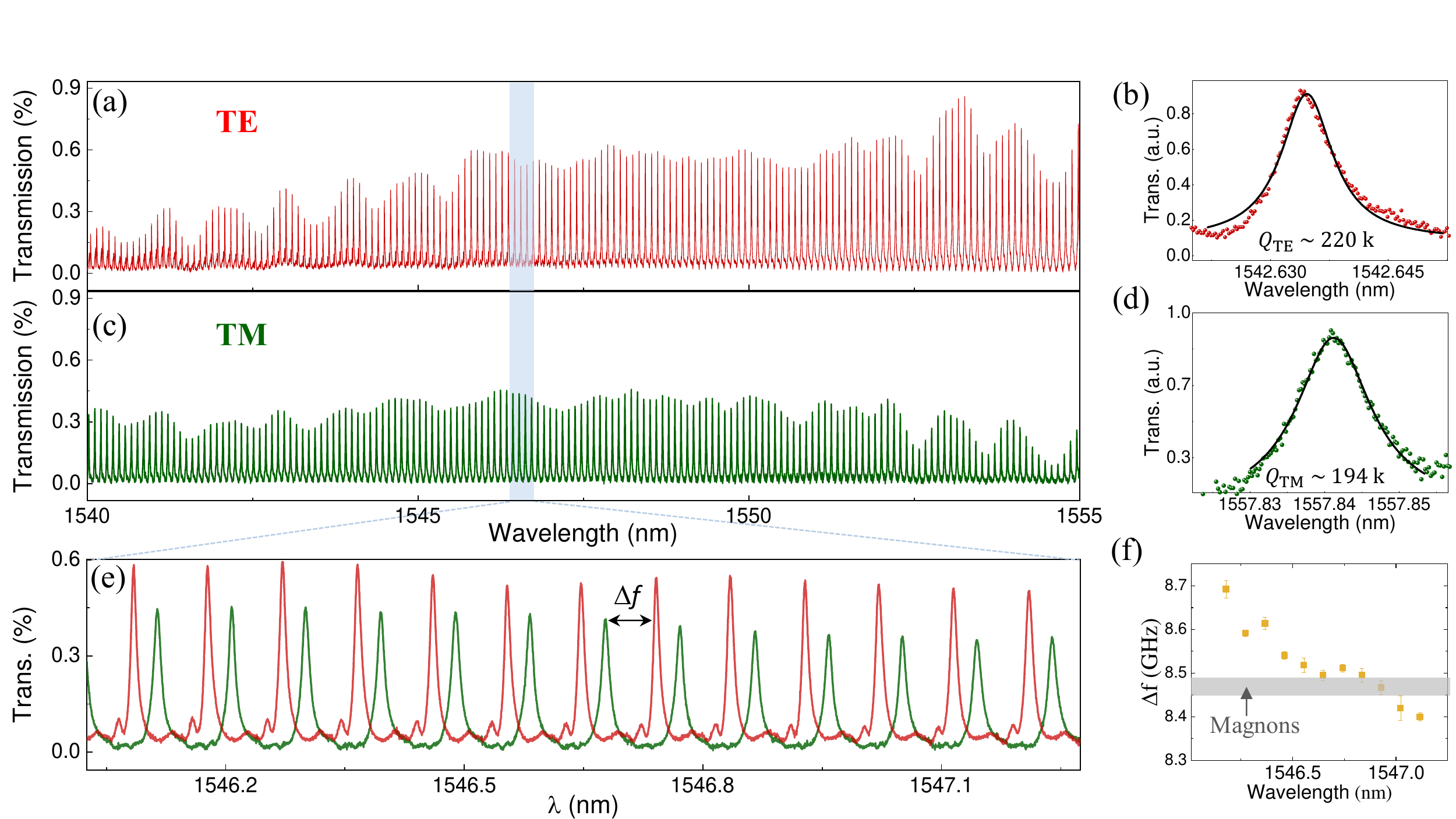}
\caption{(a) \& (c) are the optical transmission spectra of both TE (red) and TM (green) polarizations. (b) and (d) are the zoom-in single resonance spectra for the two polarizations with the fitted \textit{Q} factor, respectively. (e) is the zoom-in spectra representing the frequency difference between adjacent resonances. (f) Dispersion engineering of the frequency difference between two neighboring modes, plotted against the wavelength. The grey area depicts the target frequency band where magnons are excited.}
\label{figure2}
\end{figure*}

The linear interaction system Hamiltonian of coupled modes under the rotating-wave approximation can be written as 
\begin{equation}
H_{\mathrm{int}}={\sum_{i}^{}}{\hbar}g_{\mathrm{o,}i}\left(ab^{\dagger}+a^{\dagger}b\right)\left(m_i+m_i^{\dagger}\right)+{\sum_{i}^{}}{\hbar}g_{\mathrm{e,}i}\left(cm_i^{\dagger}+c^{\dagger}m_i\right)
\label{eq:refname1}
\end{equation}
where $a\left(a^{\dagger}\right),b\left(b^{\dagger}\right),c\left(c^{\dagger}\right),m_i\left(m_i^{\dagger}\right)$ are the annihilation (creation) operators for optical pump mode, optical signal mode, microwave mode, and \emph{i}-th order magnon mode, respectively. $g_{\mathrm{e,}i}$, and $g_{\mathrm{o,}i}$ are the electro-magnonic and photon-number-enhanced magneto-optical coupling strength for the \emph{i}-th order magnon mode, respectively \cite{hill2012coherent,kusminskiy2016coupled,kusminskiy2019cavity}. 

The triple-resonance enhanced conversion process has been proposed recently \cite{zhang2016optomagnonic,haigh2016triple, osada2016cavity}, showing a dramatic enhancement of the magneto-optical coupling strength by several orders of magnitude under the realization of energy, spin, and orbit angular momentum conservation relations. In our system, the device is engineered to satisfy the triple-resonance condition, as depicted in the Fig. \ref{figure1}(b). The dispersion relations of optical transverse-electric (TE) and transverse-magnetic (TM) modes are carefully engineered to make their effective refractive indices differ slightly from each other, thus, leading to the smooth tuning of the frequency difference between two adjacent modes over a large optical scanning range. At the same time, the frequencies of magnon modes can be tuned linearly by varying the static bias magnetic field $\vec{B_\mathrm{o}}$, according to the dispersion relation $\omega_{m}\propto\gamma\left|\vec{B_\mathrm{o}}\right|$, with $\gamma = 2\pi \times 2.8 {\text{ MHz/Oe}}$ being the gyromagnetic ratio. Thus, with the magnon tunability and optical mode dispersion engineering, the frequency difference between adjacent TE and TM optical modes can be conveniently chosen to match the magnon resonant frequencies, as well as the resonant frequency of the microwave cavity. The implementation of the microwave resonator, instead of using broadband microstrip antenna, also boosts up the magnon read-out efficiency via the cavity enhancement. Besides the engineered dispersions, the magneto-optical coupling strength is greatly enhanced by the large overlap between the optical and the magnonic modes, as shown in the mode profiles in the Fig. \ref{figure1}(b), thanks to the high mode confinement from the rib waveguide geometry.

During the experiment, a strong coherent drive tone is applied to the TE-polarized optical pump mode $\left(a\right)$, leading to a pump enhanced magneto-optical coupling strength $g_{\mathrm{o,}i}=\sqrt[]{n_a}G_{\mathrm{o,}i}$, where $n_a$ is the intracavity photon number and $G_{\mathrm{o,}i}$ is the single photon magneto-optical coupling strength. The magnons are excited by the microwave photons, due to the magnon and microwave coupling in the hybrid system. At the same time, thanks to the magneto-optical coupling, input optical photons can be inelastically scattered into a single sideband optical signal mode $\left(b\right)$ by the magnons. The on-chip conversion efficiency for the \emph{i}-th order magnon mode $\eta_i$, under the triple resonance enhanced condition, is written as (See Supplementary Material Section 4)

\begin{equation}
\eta_{i}=4\frac{C_{\mathrm{om,}i}C_{\mathrm{em,}i}\zeta_{\mathrm{e}}\zeta_{\mathrm{o}}}{{\left|1+C_{\mathrm{om,}i}+C_{\mathrm{em,}i}\right|}^2},
\label{eq:refname2}
\end{equation}
where $C_{\textrm{em,}i}\equiv$ $\frac{4{g}_{\textrm{e,}i}^2}{\kappa_{\mathrm{e}}\kappa_{\mathrm{m,}i}}$ and $C_{\textrm{om,}i}\equiv$ $\frac{4{g}_{\textrm{o,}i}^2}{\kappa_{\mathrm{o}}\kappa_{{m,i}}}$ are the electro-magnonic and magneto-optical  cooperativities, respectively, $\zeta_{\mathrm{e}}$ = $\kappa_{\mathrm{e,e}}$/$\kappa_{\mathrm{e}}$ and $\zeta_{\mathrm{o}}$ = $\kappa_{\mathrm{o,e}}$/$\kappa_{\mathrm{o}}$ are the extraction ratios. And $\kappa_{\mathrm{e,e}}$, $\kappa_{\mathrm{e}}$, $\kappa_{\mathrm{o,e}}$ and $\kappa_{\mathrm{o}}$ are the external coupling and total dissipation rates for microwave and optical signal modes, respectively. $\kappa_{{m,i}}$ is the \emph{i}-th magnon dissipation rate. It is worth noting that the total on-chip conversion efficiency $\eta$ has multiple contributions from adjacent magnon modes ($\eta \geq \eta_{i}$), and $\eta \approx \eta_{i}$ is only valid when $\kappa_{\mathrm{e}}$ and $g_{\mathrm{e},i}$ are much smaller than the FSR of the magnon modes.

\begin{figure*}[htbp]
\centering
\includegraphics[width= 147 mm]{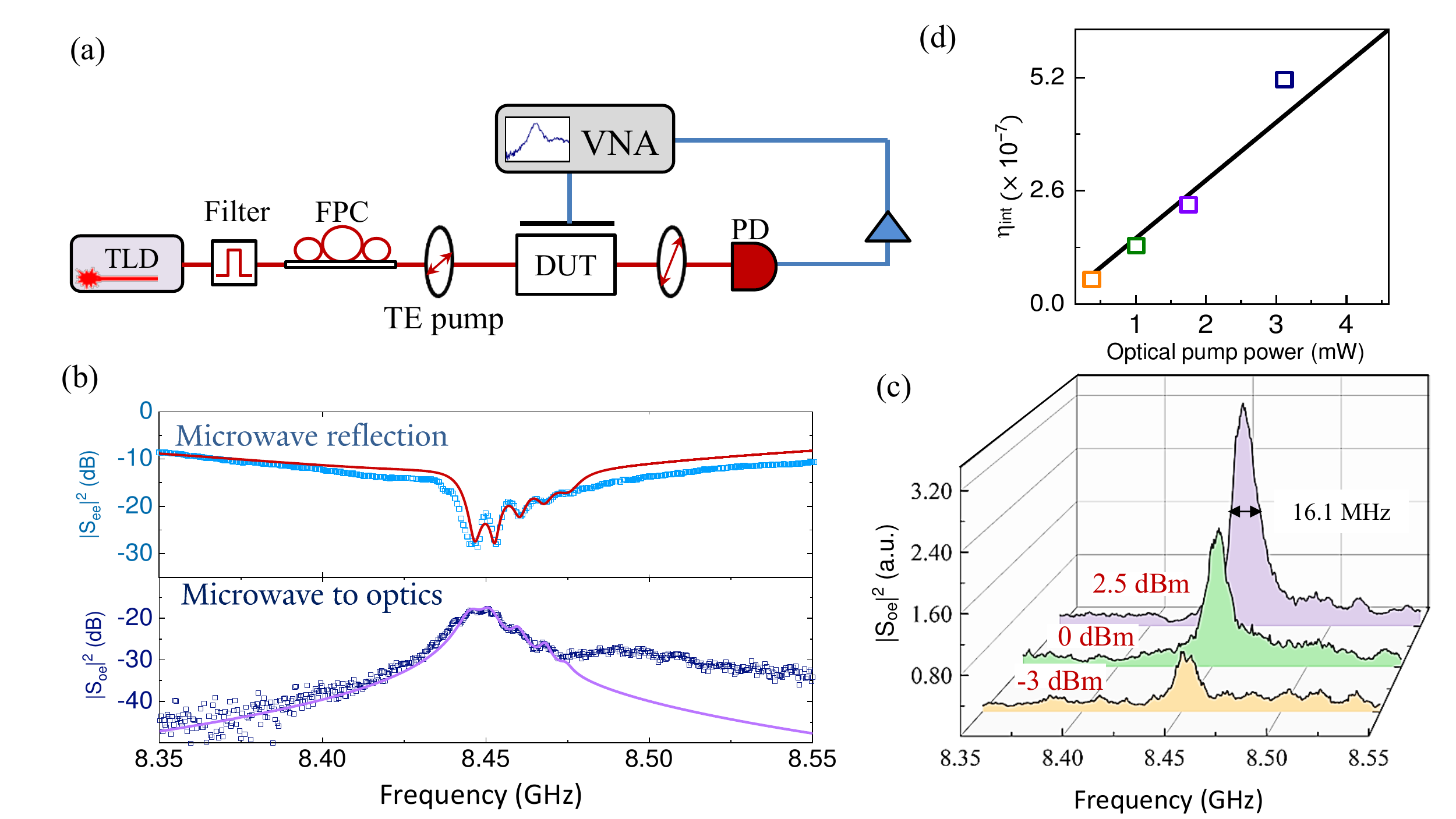}
\caption{(a) Illustration of the experimental setup. The polarization of the light from a tunable laser diode (TLD) is adjusted by a fiber polarization controller (FPC) and passes through a polarizer aligned at TE polarization. The microwave input is sent into the device-under-test (DUT) via a vector network analyzer (VNA). The beat signal between the optical pump and the converted sideband field is measured using a polarizer and a fast photodetector (PD), and fed into VNA after the signal amplification. (b) The measured (scatter) and fitted (line) microwave reflection spectrum $|S_{\mathrm{ee}}|^2$, and the corresponding microwave-to-optical conversion spectrum $|S_{\mathrm{oe}}|^2$. (c) The optical power dependence of the microwave-to-optical conversion magnitude $|S_{\mathrm{oe}}|^2$. The conversion bandwidth at full width of half maximum (FWHM) is around 16.1 MHz with 2.5 dBm pump power. (d)  Extracted (scatter) and fitted (line) internal conversion efficiency as a function of the optical pump power. The orange, green, purple, and navy squares correspond to the colored curves in (b) \& (c) respectively.}
\label{figure3}
\end{figure*}


\section{DEVICE DESIGN AND  TRIPLE-RESONANCE INTEGRATION}
A rib waveguide is fabricated in a 5-$\mu$m-thick single crystalline YIG thin film (<111> oriented) on the 500-$\mu$m-thick Gadolium Gallium Garnet (GGG) substrate that simultaneously supports optical and magnon resonances. The top layer of the rib waveguide has a width of 5 µm and a height of 1.2 µm, as shown in the Fig. \ref{figure1}(c), which is superimposed on a 200-µm-wide bottom rib layer along its center line. The bottom layer of the YIG rib waveguide, serving as a magnonic waveguide, is etched by 5 µm into the GGG layer for the spin wave mode confinement. The waveguide is 4.2-mm long, confining spin waves in both longitudinal and transverse directions. The magnetic field distribution of the magnon modes at the cross-section view is plotted in the Fig. \ref{figure1}(b). The surface of the device is covered by a 6-µm-thick silicon dioxide layer for mechanical protection and minimizing the optical scattering loss. The optical light is sent into the waveguide via a cleaved optical fiber. A copper coplanar half-$\lambda$ microwave resonator is placed beneath the YIG waveguide, which can be used to excite magnons efficiently. With high-precision fabrication and metallic reflection coating on the highly polished waveguide facets, the YIG waveguide Fabry-P\'erot cavity, which is inherently an excellent magnonic cavity, exhibits high optical quality ($Q$) factors for both TE and TM polarizations with engineered dispersion relations to support the triple resonance condition.

The optical waveguide Fabry-P\'erot cavity supports both TE and TM fundamental optical modes with the slight difference in their effective refractive indices, resulting in two sets of modes with very close \emph{free spectral range} (FSR). Here, the silicon dioxide cladding layer, YIG layer, and GGG substrate have the refractive indices of 1.44, 2.20, and 1.94, respectively. The field distributions of the fundamental optical modes (TE \& TM) are confined in the waveguide center, where the 5-$\mu$m-wide etched step locates. As we can see from the optical mode profiles in  Fig. \ref{figure1}(b), the field distributions are very similar for both TE and TM, with an aspect ratio around 1 and a mode size $\sim$ 5 $\times$ 5 ${\mu{\textrm{m}}}^2$. At the same time, both the height and width of the rib waveguide are larger than the optical wavelength ($\sim$ 1.55 $\mu$m), resulting in relatively small geometry introduced dispersion. The effective refractive indices of TE and TM modes are simulated via COMSOL Multiphysics with the values to be 2.1937 and 2.1934, respectively, close to the bulk YIG refractive index. With a small dispersion difference, the frequency difference between adjacent TE and TM optical modes slowly varies within the measured wavelength range, which can conveniently match input microwave frequencies within a single device.

Figures \ref{figure2}(a) \& (c) show the optical transmission spectra for both TE and TM input lights. The optical \emph{Q} factors for both polarizations are fitted in Figs. \ref{figure2}(b) \& (d), with a value achieving near 200,000 for both polarizations. As shown in Fig. \ref{figure2}(e), the TE and TM modes have very similar dispersion relations, and thus, the frequency difference between two adjacent modes can smoothly vary as a function of wavelength. The relation between ${\Delta}f$ and the wavelength is extracted in Fig. \ref{figure2}(f). The grey area denoted the frequency range where the magnon modes locate. The frequency difference between two adjacent modes smoothly varies from 8.7 GHz to 8.4 GHz in a measurement range from 1546 to 1547.3 nm, triggering triple-resonance enhanced frequency conversion when $\Delta{f}$ matches the magnon frequency.

\section{Microwave to Optical Conversion}

The experimental setup for measuring the microwave-to-optical conversion is depicted in Fig. \ref{figure3}(a). The optical pump is sent to the device from a tunable laser, with the polarization controlled by a fiber polarization controller and a polarizer. The device is biased perpendicularly by a static magnetic field with the field around 4800\,Oe. The microwave input is sent to the microstrip feedline via a VNA to excite the magnons. The beat signal between the optical pump field and the converted sideband field is measured using a polarizer and a fast photodiode, then fed back into the VNA.

We first characterize the magnon modes by measuring the microwave reflection spectrum. The magnon modes are confined in the slab layer of the YIG rib waveguide centered at the rib structure. Since the film is perpendicularly magnetized, the FVMSWs are excited which form standing waves between the two waveguide facets along the length direction \cite{zhang2016superstrong,chen2018strong,liu2018long}, whenever the wavevectors \emph{k} equal $i{\pi}/l$, where $l=4.2$ mm is the length of the waveguide and \emph{i} is the mode number (\emph{i} = 1, 3, 5, 7,…). The magnonic waveguide is aligned at the center of the microwave coplanar cavity, and driven by a nearly uniform \textit{rf} field, as plotted in Fig. \ref{figure1}(d). Here, only these modes with odd mode number can be excited efficiently, because the even modes have cancelled coupling strength with the uniform microwave cavity field. 

The microwave reflection spectrum is shown in  Fig. \ref{figure3}(b), illustrating the coupling between the microwave mode and multiple magnon modes. The microwave mode has the resonance around 8.444 GHz with the intrinsic and external dissipation rates $\kappa_{\mathrm{e,i}}/2\pi$ = 85 MHz and $\kappa_{\mathrm{e,e}}/2\pi$ = 165 MHz, respectively (See Supplementary Material Section 3). The multiple magnon modes are under-coupled with the microwave cavity with an FSR around 7 MHz and resonant linewidth $\kappa_{\mathrm{m,}i}/2\pi$ around 3.6 MHz for each magnon resonance. The electro-magnonic coupling strength between the fundamental magnon mode and microwave mode $g_{\mathrm{e,1}}/2\pi$ has been extracted to be 13.3 MHz, corresponding to an electro-magnonic cooperativity $C_\mathrm{em,1}$ = 4$g_\mathrm{e,1}^2$/$\kappa_{\mathrm{e}}\kappa_{\mathrm{m,1}}$ to be 0.80. Up to four magnon modes are clearly observed in the reflection spectrum. When the optical pump mode is at 1547.021 nm, according to the transmission spectra, the frequency difference between optical pump and signal modes is around 8.445 GHz, fulfilling the triple resonance condition for the lowest order magnon mode.

\begin{figure}[htbp]
\centering
\includegraphics[width=\linewidth]{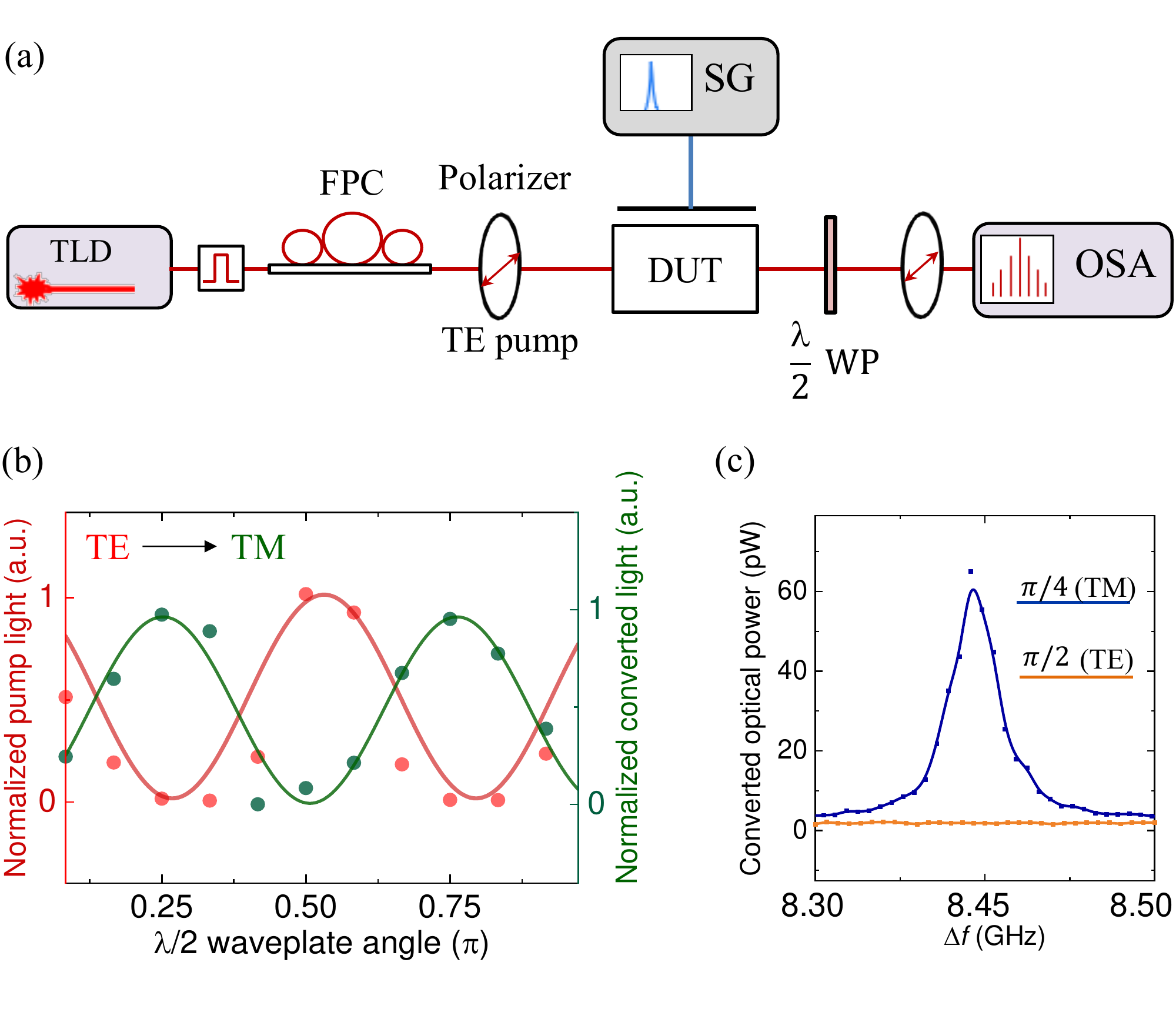}
\caption{(a) The experimental setup to measure the polarization relation of the optical pump and converted lights. The signal generator (SG) is used for the microwave input. The ${\lambda}/{2}$ waveplate (WP) is used to adjust the polarizations of the output lights, then sent through a polarizer to a optical spectrum analyzer (OSA). (b) The orthogonal optical polarizations between the pump light and the converted light. The solid lines are the sinusoidal fitting. (c) Power of the converted sideband when the waveplate is rotated at different angles.}
\label{figure4}
\end{figure}

\begin{table*}[htbp]
\centering
\caption{\bf List of parameters}
\begin{tabular}{ccccccccc}
\hline
$i$-th magnon mode & $\frac{\omega_{m,i}}{2\pi}$ (GHz) & $\frac{\kappa_{m,i}}{2\pi}$ (MHz) & $\frac{g_{\mathrm{e,}i}}{2\pi}$ (MHz) & $C_{\mathrm{em},i}$ & $\frac{g_{\mathrm{o},i}}{2\pi}$ (kHz) & $C_{\mathrm{om},i}$ ($\times10^{-7}$) & $\frac{G_{\mathrm{o},i}}{2\pi}$ (Hz) & $\eta_{i}$ ($\times10^{-7}$)\\
\hline
$i = 1$ & 8.445 & 3.55 & 13.3 & 0.80 & 22.59 & 3.72 & 17.2 & 3.68\\
$i = 3$ & 8.451 & 3.25 & 13.2 & 0.87 & 22.59 & 4.06 & 17.2 & 4.04\\
$i = 5$ & 8.460 & 3.50 & 8.5 & 0.33 & 14.68 & 1.59 & 11.18 & 1.17\\
$i = 7$ & 8.468 & 3.22 & 5.0 & 0.12 & 12.42 & 1.21 & 9.46 & 0.45\\
\hline
\end{tabular}
  \label{t1}
\end{table*}

The microwave-to-optical conversion spectrum $|S_{\mathrm{oe}}|^2$ is measured by injecting a microwave field ($-3$ dBm) and monitoring the optical beating signal, which is simultaneously taken with the reflection spectrum of $|S_{\mathrm{ee}}|^2$ in the Fig. \ref{figure3}(b) upper panel. The results show the broad conversion band peaked near the lowest order magnon mode frequency with the detectable frequency span broader than 50 MHz, thanks to the multi-mode assisted frequency conversion process, which is two to three orders of magnitude broader than other conversion devices \cite{andrews2014bidirectional}. 

The system conversion efficiency is calibrated (Supplementary Material Section 4), and the highest on-chip efficiency is achieved at the third order ($i$ $=$ 3) magnon mode, estimated to be ${1.08 \times 10^{-8}}$ at the resonant frequency of 8.451 GHz when the optical pump power is 4.8 dBm. The internal conversion efficiency at the this magnon resonant frequency $\eta_{\mathrm{int}}=\eta/\zeta_{\mathrm{e}}\zeta_{\mathrm{o}}$ is calibrated to be ${5.19 \times 10^{-7}}$. From the measurement results, the single photon magneto-optical coupling strength is fitted to be $G_\mathrm{o,3}/2\pi$ = 17.2 Hz (Supplementary Material Section 4). Such coupling strength is four orders of magnitude higher than the frequency conversion realized at YIG bulk crystal \cite{hisatomi2016bidirectional}, and 50 times higher than the result demonstrated at the YIG sphere utilizing WGM-resonance-enhancement \cite{zhang2016optomagnonic,PhysRevLett.121.199901,osada2016cavity}. With 4.8 dBm pump light, yielding 1.77 $\times 10^{6}$ intra-cavity photon number, the photon-number-enhanced coupling strength $g_\mathrm{o,3}/2\pi$ = 22.59 kHz, corresponding to an enhanced magneto-optical cooperativity $C_\mathrm{om,3}$ = 4$g_\mathrm{o,3}^2$/$\kappa_{\mathrm{o}}\kappa_{\mathrm{m,1}}$ = ${4.06 \times 10^{-7}}$. The parameters for the first four magnon modes are fitted and listed in the Table \ref{t1}. The $|S_{\mathrm{oe}}|^2$ spectrum is also fitted by considering the frequency conversion assisted by the first four magnon modes ($i$ $=$ 1, 3, 5, 7), as shown in the solid line of the Fig. \ref{figure3}(b) bottom panel. The fitting results matches very well with the measurement up to the resonant frequency of 7-th magnon mode ($i$ $=$ 7, $\sim$ 8.47 GHz). Here, the discrepancy between the fitting and measurement at high frequency tail end is due to the conversion process assisted by the higher order magnon modes along both longitudinal and transverse directions, which are not resolvable in the reflection $|S_{\mathrm{ee}}|^2$ spectrum, and thus, not included in the fitting.

The $|S_{\mathrm{oe}}|^2$ spectra yield the broadening effect instead of discrete magnon features as shown in the reflection spectrum. This is because the microwave-to-optical conversion is non-zero for the other adjacent magnon modes with the detuned frequencies. When the system satisfies the triple-resonance condition at one magnon mode, for example, the lowest-order magnon mode ($i$ = 1), the conversion process is dominated by this magnon mode due to the cavity enhancement, meanwhile, the adjacent magnon modes ($i$ = 3, 5) also contribute to the conversion. This is because the FSR of the magnon modes are relatively small ($\sim$ 7 MHz) and comparable to the magnon linewidth ($\sim$ 3.6 MHz). At the same time, the optical linewidth ($\sim$ 1.6 GHz) is much larger than the frequency band where magnons exist. Considering those factors, the adjacent modes, although are slightly detuned from the triple resonance condition, still have non-negligible contribution to the conversion process. Thus, the conversion at a specific frequency is the collective addition assisted by multiple magnon modes, enabling the broadband conversion. The detailed numerical fitting for each magnon mode and the collective conversion process is further illustrated in the Supplementary Material.

Figure \ref{figure3}(c) plots the  microwave-to-optical conversion spectra $|S_{\mathrm{oe}}|^2$ when the optical pump power is increased. The conversion efficiency increases linearly with the pump power. The 3-dB conversion bandwidth is measured to be around 16.1 MHz. In Fig. \ref{figure3}(d), the internal conversion efficiency at different optical pump power is extracted when the microwave input frequency is fixed at the lowest magnon mode with a $-$3 dBm input power, which clearly shows a linear power dependence and hence a linear magnon-to-photon conversion process.

Another distinct property of magnon-mediated microwave-to-optical conversion is the orthogonal polarizations between the pump light and converted signal light, as required by spin momentum conservation \cite{zhang2016optomagnonic,hisatomi2016bidirectional}. This feature is experimentally confirmed by tuning the polarization of both the transmitted optical pump and the converted signal light by using a ${{\lambda}/{2}}$ waveplate, as illustrated in Fig. \ref{figure4}(a). The input optical pump is aligned at TE polarization by using the fiber polarization controller and a polarizer before the device. The ${{\lambda}/{2}}$ waveplate rotates the polarization axis of both transmitted pump and the converted light together, before passing through the second polarizer for detection. The polarizer at the output side has the polarization axis aligned at TE as well. The amplitude change will behave out-of-phase, if the lights have orthogonal polarizations, which is measured by an OSA. For a ${{\lambda}/{2}}$ waveplate, if the waveplate is rotated by $\psi$, it is equivalent to rotate the polarization axis of a linear polarized light passing through it by $2\psi$ \cite{darsht1995adjustable}. As depicted in Fig. \ref{figure4}(b), the transmitted power of the pump light and signal light has $\pi⁄4$ phase difference by rotating the angle of the waveplate, corresponding to a $\pi⁄2$ angle between the polarization axis, clearly demonstrating the orthogonal polarizations between those two lights. 

We further demonstrate such polarization dependence by alternating the rotating angle $\psi$ of the ${{\lambda}/{2}}$ waveplate before the spectrometer. The converted optical sideband is measured directly via a tunable Fabry-P\'erot spectrometer. The linewidths of the measured sideband signals measured in the Fig. \ref{figure4}(c) are not the physical linewidths of the light; but instead they only represent the finite resolution (67 MHz) of the filter in the spectrometer. In this measurement, the polarizer at the output end is aligned at TE, thus, the TE-polarized light can transmit with minimal loss ($\sim$ 1 dB), but $\sim$ 32 dB rejection ratio for the TM-polarized light. When the waveplate is rotated by $\pi⁄4$, the polarization axes of both pump and converted lights are rotated by $\pi⁄2$ (TE --> TM, TM --> TE). The magnitude of the converted light is maximized at this rotating angle, while minimized when  $\psi$ $=$ $\pi⁄2$ (maintaining the original polarizations), as shown in the Fig. \ref{figure4}(c). The results verify that the converted optical sideband is TM-polarized, orthogonal to that of the pump light.

\section{Discussion and outlook}

The system conversion efficiency can be further improved by a variety of efforts in both magnonic and photonic engineering. First, the optical metallic reflective coatings can be further improved by utilizing the distributed Bragg reflector (DBR) coating on the facets to achieve higher reflectivity, leading to higher intrinsic optical \emph{Q} factors. Second, by further decreasing the device volume, such as fabricating micro-disk or ring resonators \cite{zhu2017patterned} using sputtered crystallized films \cite{zhu2017patterned}, the magneto-optical coupling strength will be dramatically improved because of further reduction of the mode volume. Third, other magnetic materials with larger Faraday constant such as ion doped YIG will offer stronger magnon-photon interaction \cite{liu2016optomagnonics}, leading to enhanced coupling strength. Lastly, measurement at cryogenic temperature will be beneficial to the improvement of the performance of microwave cavity made of superconducting materials \cite{fan2018superconducting,han2020cavity}.

In conclusion, we have developed a waveguide cavity optomagnonic system that achieves  multi-mode assisted broadband microwave-to-optical frequency conversion. By carefully engineering the optical dispersions and optimizing the mode overlaps between optical and magnon modes, the vacuum magneto-optical coupling strength has been increased by three order of magnitude compared with previous studies. In particular, a broad conversion bandwidth up to 16.1 MHz centered at 8.45 GHz has been achieved, thanks to the collective conversion process assisted by multiple magnon modes. Lastly, this optomagnonic device demonstrates high tunability for both optical and magnon modes to accommodate different microwave frequencies within one integrated device.

\section*{Funding}
We acknowledge funding from National Science Foundation (EFMA-1741666). H.X.T acknolwedge support from a DARPA MTO/MESO grant ( N66001-11-1-4114), an ARO  grant  (W911NF-18-1-0020) and Packard Foundation. 

\section*{Acknowledgments}
The authors acknowledge fruitful discussions with the team members of the EFRI Newlaw program led by Ohio State University: Andrew Franson, Seth Kurfman, Denis R. Candido, Katherine E. Nygren, Yueguang Shi, Kwangyul Hu, Kristen S. Buchanan, Michael E. Flatté,
and Ezekiel Johnston-Halperin. The authors thank M. Power and M. Rooks for the assistance in device fabrications.

\medskip


\medskip


\section*{Supplemental Documents}
See Supplement 1 for supporting content.


\bibliography{reference}

\bibliographyfullrefs{reference}

\end{document}